\documentclass[superscriptaddress,twocolumn,showpacs,preprintnumbers,amsmath,amssymb]{revtex4}

\usepackage{graphicx}
\usepackage{dcolumn}
\usepackage{bm}
\usepackage{amsmath}
\usepackage{hyperref}
\usepackage{dsfont}

\begin{document}

\preprint{APS/123-QED}

\title{Spin-squeezing and Dicke state preparation\\ by heterodyne measurement}

\author{T.~Vanderbruggen}
\author{S.~Bernon}
\author{A.~Bertoldi}
\affiliation{Laboratoire Charles Fabry de l'Institut d'Optique, CNRS and Univ. Paris-Sud,\\ Campus Polytechnique, RD 128, F-91127 Palaiseau cedex, France}
\author{A.~Landragin}
\affiliation{LNE-SYRTE, Observatoire de Paris, CNRS and UPMC,\\ 61 avenue de l'Observatoire, F-75014 Paris, France}
\author{P.~Bouyer}
\affiliation{Laboratoire Charles Fabry de l'Institut d'Optique, CNRS and Univ. Paris-Sud,\\ Campus Polytechnique, RD 128, F-91127 Palaiseau cedex, France}

\date{\today}

\begin{abstract}
We investigate the quantum non-demolition (QND) measurement of an atomic population based on a heterodyne detection and show that the induced back-action allows to prepare both spin-squeezed and Dicke states. We use a wavevector formalism to describe the stochastic process of the measurement and the associated atomic evolution. Analytical formulas of the atomic distribution momenta are derived in the weak coupling regime both for short and long time behavior, and they are in good agreement with those obtained by a Monte-Carlo simulation. The experimental implementation of the proposed heterodyne detection scheme is discussed. The role played in the squeezing process by the spontaneous emission is considered. 
\end{abstract}


\maketitle


\section{Introduction}

Generation of entangled collective spin states such as spin-squeezed states (SSS) \cite{kitagawa93, wineland92} and Dicke states \cite{dicke54,stockton04,thiel07} is of first importance for quantum information \cite{katz06} and quantum metrology \cite{bouyer97,appel08, leroux10, takano09}. A non-linear evolution of the system is required to create collective entanglement: common approaches exploit interactions between the particles \cite{esteve08, gross10}, probing with squeezed light \cite{kuzmich97} or cavity mediated interaction \cite{leroux10}. The strong non-linearity of a measurement determined by the state renormalization can also lead to such states \cite{bouchoule02, kuzmich98, kuzmich00, nielsen09}.

Measurement induced squeezing of atomic ensembles has been performed using either a Mach-Zehnder interferometer and simultaneous interaction with several optical frequencies \cite{saffman09}, or cavity transmission modulation \cite{schleier10, leroux10bis}. Nevertheless, the experimental realization of highly engineered atomic states is still challenging and a deep understanding of the underlying stochastic process of the measurement is needed. The dynamic of the atomic state collapse can be described as a succession of partial measurements. Several methods have been used to model this collapse process, like stochastic Schr{\"o}dinger equation \cite{stockton04,nielsen08}, master equation \cite{sorensen02}, or wavevector formalism \cite{bouchoule02}.

In this article, we propose to use the frequency modulation (FM) spectroscopy technique \cite{bjorklund83,savalli99,lye03,teper08} to generate both spin-squeezed and Dicke states. A laser beam is phase modulated to produce frequency sidebands; one sideband is placed close to an atomic transition and experiences a phase-shift passing through the atomic sample. The detection of the beatnote at the modulation frequency allows then to estimate the atomic population of the probed state.

A theorical analysis of the measurement process shows that a heterodyne detection performed on an atomic sample allows to prepare new quantum states. The study is based on the wavevector formalism previously introduced for homodyne detection \cite{bouchoule02} and shows similarities with the method used in \cite{brune90, guerlin07} to reconstruct the photonic state from the results of a QND measurement sequence. We study the quantum trajectories of the atomic state caused by the repeated interaction with single-photons sent in the QND apparatus. In the weak coupling limit an analytical expression is derived for the evolution of the state variance during the measurement, it completely describes the atomic wavefunction collapse. The result is in good agreement with a Monte-Carlo simulation of the measurement process.

\section{The measurement process}

\begin{figure}[!h]
\begin{center}
\includegraphics[width=8cm,keepaspectratio]{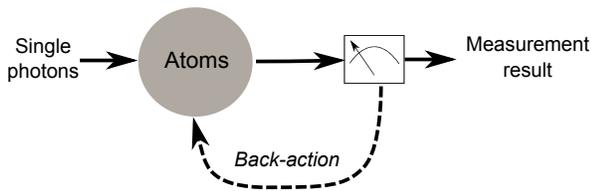}
\caption{Basic scheme of the measurement process: the phase-shift induced by the atomic sample is estimated using the repeated interaction with single-photons. The back-action on the atomic state increases with the precision of the phase measurement.}
\label{fig:measu_backaction}
\end{center}
\end{figure}

The principle of the QND measurement is presented in Fig. ~\ref{fig:measu_backaction}. An atomic system interacts coherently with an optical probe. As a result, the orientation component of the atomic state (such as the atom number in a well defined hyperfine level), becomes weakly entangled with the probe photons, and the photodetection gradually provides the observer with information about the spin state. Continuous observation conditionally reduces the orientation uncertainty with respect to the QND measurement outcome thanks to the measurement quantum back-action. In any individual measurement trajectory, the initial wavefunction will thus collapse in a single squeezed state whereas the ensemble averaged uncertainty will not be reduced, because the measurement outcomes over many QND trajectories are distributed within the variance of the initial state. Our model will allow to study the dynamics of this collapse by considering single photon detection events.

\subsection{Heterodyne QND detection}

We consider the QND measurement of the population difference between two atomic levels using an heterodyne detection. The measurement is realised by estimating the phase-shift between the two frequency components of the detection beam. This configuration is equivalent to a frequency space interferometer in which each path corresponds to a single frequency mode. In our model, we consider photons individually sent into this interferometer and reconstruct the output beatnote photon after photon by comparing their arrival time with a time reference.

\subsubsection{Model of the apparatus}

\begin{figure}[!h]
\begin{center}
\includegraphics[width=8cm,keepaspectratio]{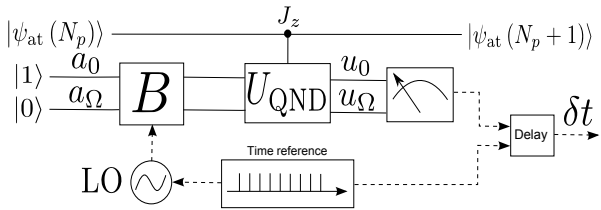}
\caption{Model for the QND based heterodyne detection. Single-photons are sent through a spectral beamsplitter $B$ and interact non-destructively with the atomic ensemble ($U_{\rm QND}$). The detection time of each photon is measured by comparison with a time reference signal.}
\label{fig:het_det_scheme}
\end{center}
\end{figure}

The model for the QND apparatus is shown in Fig.~\ref{fig:het_det_scheme}. The input optical state is composed of two spectral modes. The probe mode $(0)$ at a frequency $\omega_{0} / 2 \pi$ is close to the frequency of the atomic transition, and will therefore be strongly phase shifted when traveling through the atoms. The reference mode ($\Omega$) is at a far-off resonance frequency $\left( \omega_{0} + \Omega \right) / 2 \pi$ and will be unaffected. A single-photon state $\left| 1 \right\rangle$ in the mode (0) is associated to the photon annihilation operator $a_{0}$, and the vacuum state $\left| 0 \right\rangle$ in the mode ($\Omega$) is associated to $a_{\Omega}$. A  single sideband optical phase modulator $B$ is used as a spectral beamsplitter to generate spectrally mode-entangled single-photons (see App.~\ref{app:sp_bs}). A local oscillator (LO) drives $B$ with frequency $\Omega / 2 \pi$. We consider the driving signal as a noise-free classical field. After passing through the atoms, the two-channel-optical-state is sent to a single-photon detector which is assumed to be ideal. The detection of a photon in the mode (0) is associated with the operator $u_{0}$, whereas the detection of a photon in the mode ($\Omega$) is associated with $u_\Omega$.  If the coherence length of the optical source is much larger than the LO wavelength, the two modes (0) and  ($\Omega$) are in a coherent superposition, and the spatial overlap of these two modes generates a beating at frequency $\Omega/ 2 \pi$. A pulse generator, in phase with the LO, delivers regularly spaced pulses (at time interval $\tau=2 \pi / \Omega$) used as a reference to measure the arrival time of the photons. The histogram representation of the photon arrival time allows to reconstruct the beating signal.

\subsubsection{Atomic state}

The atomic ensemble is a collection of a fixed number $N_{\mathrm{at}}$ of two-levels atoms 
$\left\{ \left| a_{i} \right\rangle , \left| b_{i} \right\rangle \right\}$, described as fictitious 1/2-spins $\{ j_{x}^{(i)} , j_{y}^{(i)} , j_{z}^{(i)} \}$ where:
\begin{eqnarray}
j_{x}^{(i)} & = & \frac{1}{2} \left( \left| b_{i} \right\rangle \left\langle a_{i} \right| + \left| a_{i} \right\rangle \left\langle b_{i} \right| \right), \\
j_{y}^{(i)} & = & \frac{i}{2} \left( \left| a_{i} \right\rangle \left\langle b_{i} \right| - \left| b_{i} \right\rangle \left\langle a_{i} \right| \right), \\
j_{z}^{(i)} & = & \frac{1}{2} \left( \left| b_{i} \right\rangle \left\langle b_{i} \right| - \left| a_{i} \right\rangle \left\langle a_{i} \right| \right).
\end{eqnarray}
The atomic ensemble is characterized by the collective spin operators $J_{k} = \sum_{i} j_{k}^{(i)}, \; k=x,y,z$. A collective state $\left| n \right\rangle$ is an eigenstate of both $J_{z}$ and $\textbf{J}^{2} = J_{x}^{2} + J_{y}^{2} + J_{z}^{2}$, and is called a Dicke state. It verifies:
\begin{eqnarray}
\textbf{J}^{2} \left| n \right\rangle & = & \frac{N_{\rm at}}{2} \left( \frac{N_{\rm at}}{2} + 1 \right) \left| n \right\rangle \\
J_{z} \left| n \right\rangle & = & n \left| n \right\rangle
\end{eqnarray}
where $-N_{\rm at}/2 \leq n \leq N_{\rm at}/2$. After $N_{p}$ photons detected the atomic state is $\left| \psi_{\rm at} \left( N_{p} \right) \right\rangle = \sum_{n} c_{n} \left( N_{p} \right) \left| n \right\rangle$. The initial atomic state is a coherent spin state (CSS) polarized along $J_{x}$, which means an average population difference $\left\langle J_{z} \right\rangle = 0$, and its explicit expression is \cite{zare}:
\begin{equation}
c_{n} \left( 0 \right) = \frac{1}{2^{N_{\mathrm{at}}/2}} \sqrt{\frac{N_{\rm at}!}{\left( \frac{N_{\rm at}}{2} + n \right)! \left( \frac{N_{\rm at}}{2} - n \right)! }}.
\end{equation}
From the Moivre-Laplace theorem, if $N_{\mathrm{at}} \gg 1$ the coefficients $c_{n} \left( 0 \right)$ are well approximated by a gaussian distribution:
\begin{equation}
c_{n} \left( 0 \right) \propto \exp \left( - \frac{n^{2}}{N_{\mathrm{at}}} \right).
\label{eqn:initial_state}
\end{equation}

\subsection{Analytical description of the measurement process}

\subsubsection{Scattering matrix of the interferometer}

The spectral beamsplitter $B$ creates a coherent superposition of the two input spectral modes ($0$) and ($\Omega$), see App.~\ref{app:sp_bs}. Its action is described by the following matrix:
\begin{equation}
B \propto \left(
\begin{array}{cc}
\sqrt{T}                & -\sqrt{R}              \\
\sqrt{R} e^{i \Omega t} & \sqrt{T}e^{i \Omega t} \\
\end{array}
\right),
\end{equation}
where, in analogy with a spatial beamsplitter, $R$ is the probability for a photon in the probe mode ($0$) to be in the reference mode ($\Omega$) at the output of $B$, and $T$ is the probability for a photon to stay in the carrier mode. We neglect any absorption and dispersion in $B$ so that $R+T=1$.

When $\omega_{0}+\Omega$ is far from any atomic transition, the reference beam experiences no phase-shift and the QND interaction can be written as
\begin{equation}
U_{\mathrm{QND}} \left( J_{z} \right) = \left(
\begin{array}{cc}
e^{i \phi J_{z}} & 0 \\
0                & 1 \\
\end{array}
\right).
\end{equation}
$\phi$ is the optical phase-shift induced by a number diference of two ($n=1$), and depends on the coupling strength of the transition at $\omega_{0}$ and on the optical density of the atomic cloud. Spontaneous emission induced by the probe beam is neglected in this analysis; the reduction it causes on the coherence of the atomic state is considered in Sec.~\ref{sec:exp_cons}.

The scattering matrix of this system is $S\left( J_{z} \right) = U_{\mathrm{QND}}\left( J_{z} \right) B$ and it furnishes the output modes of the interferometer once the input ones are given: $\left( u_{0} \left( J_{z} \right), u_{\Omega} \left( J_{z} \right) \right) = S\left( J_{z} \right) \left( a_{0}, a_{\Omega} \right)$.

\subsubsection{Measurement back-action}

A click at the detector corresponds to the annihilition of a photon either in mode ($0$) or ($\Omega$). After the measurement the state of the system satisfies the relation:
\begin{eqnarray}
& \left| 0_{0}, 0_{\Omega} \right\rangle \otimes \left| \psi_{\rm at} \left( N_{p}+1 \right) \right\rangle \nonumber \\
& \propto  \left[ u_{0} \left( J_{z} \right) + u_{\Omega} \left( J_{z} \right) \right] \left| 1_{0}, 0_{\Omega} \right\rangle \otimes \left| \psi_{\rm at} \left( N_{p} \right) \right\rangle.
\end{eqnarray}
The stochastic recurrence relation which defines the atomic state evolution after one measurement follows from the expression of $S \left( J_{z} \right)$:
\begin{eqnarray}
\left| \psi_{\rm at} \left( N_{p}+1 \right) \right\rangle \propto 
\left[ \left( \sqrt{T} + \sqrt{R} \right) \cos \left( \frac{\phi J_{z}}{2} - \frac{\Omega \widetilde{t}}{2} \right) \right. \nonumber \\
\left. + i \left( \sqrt{T} - \sqrt{R} \right) \sin \left( \frac{\phi J_{z}}{2} - \frac{\Omega \widetilde{t}}{2} \right) \right] \left| \psi_{\rm at} \left( N_{p} \right) \right\rangle, \label{eqn:rec_relation}
\end{eqnarray}
where $\widetilde{t}$ is the time at which the photon has been detected.

This expression contains a beatnote signal at frequency $\Omega/ 2 \pi$ with a contrast $\mathcal{C} = 2 \sqrt{RT}$. The reconstruction of the beatnote photon after photon is achieved by comparing the arrival time $\widetilde{t}_{k}$ of the $k$-th photon with the last time reference pulse $p \tau$, where $p$ is the number of pulses counted between $t=0$ and $t=\widetilde{t}_{k}$. We estimate the phase-shift $\widetilde{\varphi}_{k} \equiv \Omega \widetilde{\delta t}_{k}$ from the time delay $\widetilde{\delta t}_{k}= \widetilde{t}_{k} - p \tau$ between the last pulse of the time reference and the photon detection.

The atomic state after the detection of $N_{p}$ photons is obtained applying $N_{p}$ times the relation of Eq.~(\ref{eqn:rec_relation}) on the initial atomic state $\left| \psi_{\rm at} \left( N_{p} = 0 \right) \right\rangle$. Expressed as a superposition of collective states, it results $\left| \psi_{\rm at} \left( N_{p} \right) \right\rangle = \sum_{n} \mathcal{F}_{N_{p}} \left( n \right) c_{n} \left( 0 \right) \left| n \right\rangle$, where
\begin{equation}
\left| \mathcal{F}_{N_{p}} \left( n \right) \right|^{2} \propto \prod_{k=1}^{N_{p}} \left[ 1 + \mathcal{C} \cos \left( \phi n - \widetilde{\varphi}_{k} \right) \right]
\label{eqn:F_het}
\end{equation}
is the back-action function \footnote{It is interesting to note that, in the general case, the process is \textit{non Markovian} since the phase-shift $\widetilde{\varphi}_{k}$ depends on all the previous detection results. This is a main difference between the heterodyne and the homodyne detection: in the latter the evolution is Markovian independently of the coupling regime, since only two measurement outcomes are possible \cite{bouchoule02}.}.

\begin{figure*}
\includegraphics[width=18cm,keepaspectratio]{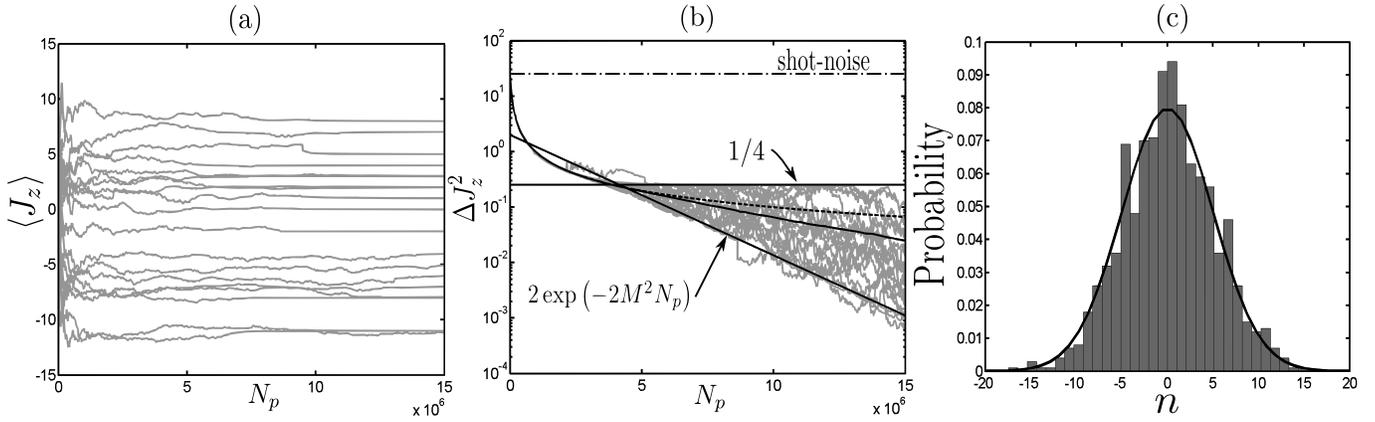}
\caption{Trajectories of the atomic distribution momenta obtained with a Monte-Carlo simulation, once set $N_{\mathrm{at}} = 200$, $\phi = 10^{-3}$ rad, and $R = T = 1/2$. (a) Mean position of the atomic population $\left\langle J_{z} \right\rangle$ as a function of the number of measured photons. For a long enough measurement the state converges to a Dicke state. (b) Variance evolution during the measurement sequence. The solid line is the average over 1000 trajectories (only 20 are plotted) while the dashed line is the result of the analytical calculation of the variance at short time, given by Eq.~(\ref{var_Jz}). At short time scale, the evolution is deterministic as predicted by the analytical study. At long time scale, the variances become stochastic but remain bounded. In addition, the average variance over the trajectories is below the short time behavior. (c) The histogram of the mean positions at the end of the trajectories is compared to the initial distribution (solid line). The measurement process satisfies the Born probability rule $P_{n} = \left| \left\langle n |\psi_{\rm at} \left( N_{p}=0 \right) \right\rangle \right|^{2}$.}
\label{fig:simul_het}
\end{figure*}

Eq.~(\ref{eqn:F_het}) leads straightforwardly to the probability $P \left( \varphi \right)$ to measure a phase-shift $\varphi$ when the $\left( N_{p}+1 \right)$-th photon is detected:
\begin{equation}
P \left( \varphi \right) = \frac{1}{2 \pi} \sum_{n} \left| c_{n} \left( N_{p} \right) \right|^{2} \left( 1 + \mathcal{C} \cos \left( \phi n - \varphi \right) \right).
\label{eqn:P_varphi}
\end{equation}
Using this expression and the recurrence relation of Eq.~(\ref{eqn:rec_relation}), we simulated the quantum trajectories followed by the atomic state (see App.~\ref{app:simul}). Typical results are presented in Fig.~\ref{fig:simul_het}.

In the weak coupling limit ($\phi N_{\rm at} \ll 1$), it is possible to derive an analytical expression for the $\left| \mathcal{F}_{N_{p}} \left( n \right) \right|^{2}$ function (see App.~\ref{app:deriv_F}):
\begin{equation}
\left| \mathcal{F}_{N_{p}} \left( n \right) \right|^{2} \propto \exp \left[ - 2 M^{2}N_{p} \left( n^{2} + 2 \overline{\delta \widetilde{\varphi}} n /\phi \right) \right],
\label{eqn:F_gauss}
\end{equation}
where 
\begin{equation}
\overline{\delta \widetilde{\varphi}} = \lim_{\frac{N_{t}}{N_{p}} \rightarrow 0} \frac{N_{t}}{N_{p}} \sum_{j=1}^{N_{p}/N_{t}} \delta \widetilde{\varphi}_{j}
\end{equation}
is the average mean position over the followed trajectory, and 
\begin{equation}
M^{2} = \frac{\phi^{2}}{4} \left( 1- \sqrt{1-\mathcal{C}^{2}} \right),
\end{equation}
is the measurement strength. This expression of the back-action function contains the complete description of the atomic state evolution in the weak coupling regime and allows to quantitatively study the atomic wavefunction collapse, that is the squeezing process.

\section{Dynamics of the wavefunction collapse}

The atomic state evolution is studied by comparing the initial atomic state distribution $c_{n} \left( 0 \right)$ of Eq.~(\ref{eqn:initial_state}) with $\mathcal{F}_{N_{p}} \left( n \right)$. Depending on their relative width, two regimes appear during the atomic state evolution: the short time limit - where the width of the atomic distribution $\Delta J_{z}^{2}$ is large compare to one and the atomic state is in a superposition of many atomic levels - and the long time limit - where the distribution is very narrow and the state is split on a few levels of amplitude $\mathcal{F}_{N_{p}} \left( n \right)$. The boundary between these two regimes occurs for $\Delta J_{z}^{2} \sim 1$, that is for $N_{p} \sim M^{-2}$.

\subsection{Short time limit}

In the short time limit, defined by the condition $N_{p} \ll M^{-2}$, $\left| \mathcal{F}_{N_{p}} \left( n \right) \right|^{2}$ is broad and $\left| c_{n} \left( N_{p} \right) \right|^{2}$ is a gaussian distribution characterized by the following mean position and variance:
\begin{eqnarray}
\left\langle J_{z} \right\rangle & = & - \mathcal{C}^{2} \xi^{2} \kappa^{2} \overline{ \delta \widetilde{\varphi}}/\phi, \label{mean_Jz}\\
\Delta J_{z}^{2} & = & \xi^{2} N_{\mathrm{at}}/4, \label{var_Jz}
\end{eqnarray}
where $\kappa^{2} = M^{2} N_{\mathrm{at}} N_{p}$ is the signal-to-noise ratio and $\xi^{2} = 1 / \left( 1+\kappa^{2} \right)$ is the squeezing factor. For $N_{p} \geq 1$, the squeezing factor drops below unity and the initial CSS collapses into a SSS as a consequence of the measurement process. The remarkable result is that $\Delta J_{z}^{2}$ is independent of the stochastic parameter $\overline{\delta \tilde{\varphi}}$ at first order and is thus deterministic, as shown with the numerical simulation in Fig.~\ref{fig:simul_het} (b).

\begin{figure}[!h]
\begin{center}
\includegraphics[width=8cm,keepaspectratio]{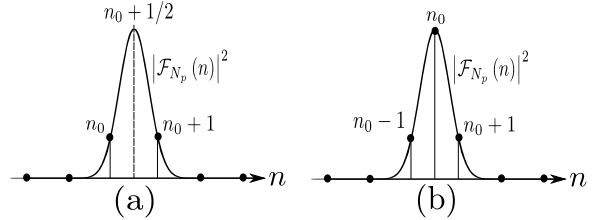}
\caption{Extreme cases for long time evolution. The mean position of the atomic distribution is in the middle of two eigenvalues of $J_{z}$, (a) or centered on an eigenvalue of $J_{z}$ (b).}
\label{fig:long_time_limit_cases}
\end{center}
\end{figure}

\subsection{Long time limit}

In the long time limit ($N_{p} \gg M^{-2}$) the atomic distribution is very narrow ($\Delta J_{z}^{2} \ll 1$) and the atomic state $c_{n} \left( N_{p} \right) \sim \mathcal{F}_{N_{p}} \left( n \right)$ is spread over a few eigenstates. The evolution strongly depends on the distance between the mean value $\left\langle J_{z} \right\rangle$ and the closest eigenvalue of $J_{z}$. Two extreme cases can be considered (Fig.~\ref{fig:long_time_limit_cases}): either $\left\langle J_{z} \right\rangle$ is in the middle of two eigenvalues, which turns out to be an unstable equilibrium state of the system, or it is an eigenvalue of $J_{z}$, which corresponds to the stable equilibrium state (the stability of the attractors in treated in \cite{stockton04,adler01}). 

In the first case the distribution is centered between two eigenstates of $J_{z}$, namely $\left| n_{0} \right\rangle$ and $\left| n_{0} +1 \right\rangle$. The atomic state is well approximated by a superposition of these two states $\left| \psi_{\rm at} \left( N_{p} \right) \right\rangle \approx \left( \left| n_{0} \right\rangle + \left| n_{0}+1 \right\rangle \right) / \sqrt{2}$, since the amplitude of the other states decreases exponentially. In this case $\left\langle J_{z} \right\rangle = n_{0}+1/2$ and the variance is $\Delta J_{z}^{2} = \left\langle \psi_{\rm at} \left( N_{p} \right) \right| J_{z}^{2} \left| \psi_{\rm at} \left( N_{p} \right) \right\rangle - \left\langle J_{z} \right\rangle^{2} = 1/4$, which results to be the upper bound value in the long time limit. This limit clearly appears in the numerical calculations, as shown in Fig.~\ref{fig:simul_het} (b).

The second case represented in Fig.~\ref{fig:long_time_limit_cases} (b) corresponds to an atomic distribution centered on an eigenstate $\left| n_{0} \right\rangle$ of $J_{z}$. The atomic state is then a superposition of three eigenstates 
\begin{equation}
\left| \psi_{\rm at} \left( N_{p} \right) \right\rangle  \approx  \sum_{k = -1,0,1} \mathcal{F}_{N_{p}} \left( n_{0}+k \right) \left| n_{0}+k \right\rangle.
\end{equation}
From the normalization condition, the symmetry of the state around $n_{0}$, and Eq.~(\ref{eqn:F_gauss}), we have:
\begin{eqnarray}
& \sum_{k = -1,0,1} \left| \mathcal{F}_{N_{p}} \left( n_{0}+k \right) \right|^{2} = 1 &\\
& \left| \mathcal{F}_{N_{p}} \left( n_{0} \pm 1 \right) \right|^{2} = \left| \mathcal{F}_{N_{p}} \left( n_{0} \right) \right|^{2} e^{- 2 M^{2} N_{p}} &.
\end{eqnarray}
Since $\left\langle J_{z} \right\rangle = n_{0}$, it follows $\Delta J_{z}^{2} = 2 \exp \left( - 2 M^{2} N_{p} \right)$, which is the lower bound for the variance evolution as plotted in Fig.~\ref{fig:simul_het} (b).

\section{\label{sec:exp_cons}Experimental considerations}

The main experimental challenge in a QND detection is to limit the spontaneous emission and so the optical power while reaching the shot-noise limit with a sufficient bandwidth. In this paragraph, we show how a heterodyne detection allows to reach the shot-noise of a weak probe beam and how the existence of two sidebands rejects all the frequency independent common dephasing sources. We finally discuss the effect of the spontaneous emission on the squeezing performances.

\begin{figure}[!h]
\begin{center}
\includegraphics[width=7cm,keepaspectratio]{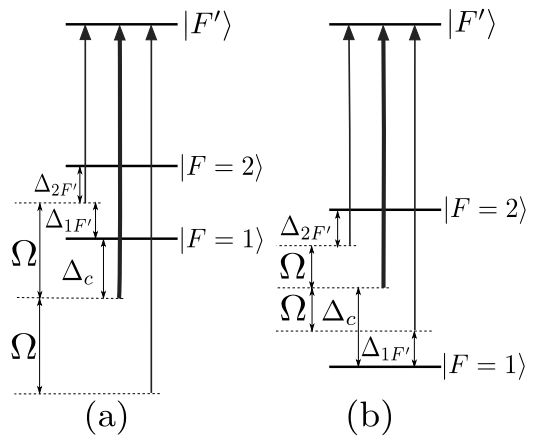}
\caption{Two possible measurement schemes. (a) One sideband is in the middle of the two probed states whereas the carrier and the other sideband are far from the transition ($\Delta_{FF'} \ll \Delta_{c}$). (b) Each sideband is close to an atomic transition, the carrier is in the middle of the two probed states.}
\label{fig:meas_strategies}
\end{center}
\end{figure}

In a realistic setup, different configurations for the measurement can be implemented depending on the relative position of the carrier and the sidebands with respect to the atomic transitions. Two examples of possible schemes are presented in Fig.~\ref{fig:meas_strategies}, where the case of alkaline atoms is considered.

\subsection{Shot-noise limited detection}

The intrinsic advantage of a heterodyne detection is that it makes possible to reach the shot-noise for a weak probe beam by beating it with a much stronger reference beam. In FM spectroscopy, the reference beam is the strong carrier and the probe beam is the weak sideband. If $N_{c}$ is the number of photons in the carrier, $N_{s}$ that one in the sideband, and $N_{e}$ the photon equivalent noise due to the detection electronics, the signal-to-noise ratio of the optical detection is:
\begin{equation}
\mathrm{SNR} \propto \sqrt{\frac{N_{c} N_{s}}{N_{c} + N_{s} + N_{e}}}.
\end{equation}
If $N_{c} \gg N_{s}, \; N_{e}$, then $\mathrm{SNR} \sim \sqrt{N_{s}}$: by using a strong reference, the detection can thus be limited to the shot-noise of the weak probe. 

Nevertheless, the reference must be far from the atomic transition to strongly reduce its absorption. A high modulation frequency $\Omega$ is thus required, which results to be the main technical limitation of the method. Nowadays, photodiodes designed for high-speed optical communications provided with integrated transimpedance amplifiers with gains of a few kilo-ohms, set an indicative upper limit of $10$ GHz for the achievable bandwidth.

\subsection{Common mode noise rejection}

All the systems where the phase induced by the atoms on a probe beam is measured by comparison with the phase of a reference beam  require a precise control of all the sources of relative phase noise between the two beams. The geometrical splitting of the two components, like in a Mach-Zehnder interferometer, requires to precisely stabilize the relative length between the two paths. On the other hand, if the splitting is in frequency and the two components are spatially overlapped, all the noise sources acting on the optical path length (like acoustic noise, optical index fluctuations and mechanical displacements) are rejected, since they are common mode. In the specific case of a two sidebands modulator, where only one sideband is close to an atomic resonance, the signal detected on the photodiode is the difference between the beatnotes of each sideband with the carrier. The differential detection leaves only the atomic contribution to the optical index, an active stabilisation of the interferometer is not required.

In \cite{saffman09}, an amplitude modulation method in a Mach-Zehnder interferometer is proposed which similarly rejects the common mode noise using two probe beams. Nevertheless, it requires a demanding good spatial overlap, a well balanced power and an active phase control of the two probes; all these features are intrinsic in the method we propose.

The proposed scheme could be further improved by injecting the probe in an optical cavity. The signal-to-noise ratio for the same destructivity is increased by a factor square-root of the finesse \cite{lye03}. The injection in another cavity resonance can be used to perform a differential measurement when studying amplitude fluctuations of a DC signal as shown in \cite{schleier10, leroux10bis}. In our scheme, a differential measurement that rejects the phase noise due to cavity length fluctuations is realized by placing the carrier in the middle of two cavity resonances and injecting the two sidebands in two different resonances.

\subsection{Spontaneous emission}

The performance of a QND detection is reduced by the spontaneous emission induced by the probe. The atomic decay causes a decoherence of the atomic state since the detection of the emitted photon allows to determine in which state the atom has been projected. The problem of quantifying this effect has been treated in \cite{bouchoule02, madsen04, echaniz05}.

We now analyse the performance of scheme (a) in Fig.~\ref{fig:meas_strategies} when the atomic decay is taken into account. The dephasing induced by the atomic cloud on the optical probe is given by \cite{oblack05}:
\begin{equation}
\Delta \Phi = \frac{\lambda^{2}}{4 \pi \mathcal{A}} \left( N_{1} S_{1} + N_{2} S_{2} \right),
\end{equation}
where $N_{F}$ is the population of the $\left|F \right\rangle$ state, $\lambda$ the optical wavelength, $\mathcal{A}$ the probe beam area, and $S_{F}$ the probe coupling to the $\left|F \right\rangle$ state:
\begin{equation}
S_{F} = \sum_{F'} \frac{\gamma \Delta_{FF'}}{\Delta_{FF'}^{2}+\gamma^{2}} S_{FF'}.
\end{equation}
In the last equation $\gamma$ is the atomic linewidth and $S_{FF'}$ is the dipole transition strength of $\left|F \right\rangle \rightarrow \left|F' \right\rangle$ given by:
\begin{equation}
S_{FF'} = \left( 2 F' + 1 \right) \left( 2 J + 1 \right) \left\{
\begin{array}{ccc}
J  & J' & 1 \\
F' & F  & I \\
\end{array}
\right\}^{2},
\end{equation}
where $I$ and $J$ are the nuclear and total angular momentum, respectively. The detunings $\Delta_{1F'}$ and $\Delta_{2F'}$ are choosen so that $S_{1} = -S_{2} \equiv S$. Introducing $\epsilon$ so that $N_{1} = N_{\rm at} \left( 1 + \epsilon \right)/2$ and $N_{2} = N_{\rm at} \left( 1 - \epsilon \right)/2$, the optical dephasing can be written as $\Delta \Phi = \rho_{0} S \epsilon$, where $\rho_{0} = \lambda^{2} N_{\rm at} / 4 \pi \mathcal{A}$ is the resonant optical density. The phase-shift $\phi$ induced by a population difference of one atom between the two atomic states is obtained for $\epsilon = 2/N_{\rm at}$:
\begin{equation}
\phi = \frac{\lambda^{2} S}{2 \pi \mathcal{A}}.
\end{equation}
The integrated probability for an atom to scatter a photon when probed by $N_{p}$ photons is, in the limit $\mathcal{C} \ll 1$:
\begin{equation}
\eta = \frac{\rho_{0}}{N_{\rm at}} \frac{\mathcal{C}^{2}N_{p}}{2} \mathcal{L},
\end{equation}
where $\mathcal{L}$ is the atomic lineshape:
\begin{equation}
\mathcal{L} = \sum_{F, F'} \frac{\gamma^{2}}{\Delta_{FF'}^{2}+\gamma^{2}} S_{FF'}.
\end{equation}

The signal-to-noise ratio can be rewritten as a function of the decoherence: $\kappa^{2} = \mu \rho_{0} \eta$ where $\mu = S^{2} / \mathcal{L}$. In \cite{madsen04}, the squeezing factor is estimated to be:
\begin{equation}
\xi^{2} = \frac{\left( 1 - \eta \right)^{2}}{1 + \mu \rho_{0} \eta} + 1 - \left( 1 - \eta \right)^{2}.
\end{equation}

Considering $10^{7}$ atoms of $^{87}$Rb optically trapped with a beam waist of $20 \; \mu$m and probed on the $D_{2}$ ($5^{2} S_{1/2} \rightarrow 5^{2} P_{3/2}$) line at $\lambda=780$ nm with a detuning $\Delta_{13}=3.2 $ GHz then $\mu \sim 1$ and $\phi = 4.1 \; 10^{-7}$ rad. The resulting resonant optical density is $\rho_{0} \sim 2400$ from which a squeezing factor $\xi^{2} = 0.06$ (about $12$ dB) can be achieved for an optimum decoherence $\eta \sim 0.01$. It corresponds to a measurement of $N_{p} \sim 10^{7}$ photons in the total beam for a modulation depth of $1$\%. 

The optimum squeezing factor strongly depends on the optical density of the atomic sample. Using a Bose-Einstein condensate (BEC), a much higher squeezing can be achieved but much less photons have to be used during the measurement to reach the optimum.

With regard to the possibility of exploring the long time regime behavior to prepare Dicke states, the decoherence due to spontaneous emission can be seen as a constraint that would make this regime hard to reach. Nevertheless the discussion above holds mainly for a large atom number, if the initial sample contains only a few atoms (for example a BEC of 100 atoms) the variance of the coherent state is of the order of a few units so that entering the Dicke state regime in the presence of spontaneous emission is realistic.

\section{Conclusion}

In this article we analyse a QND heterodyne detection of an atomic sample to prepare spin-squeezed states. A theorical model of the measurement process is proposed that consists on the sequential detection of single photons that causes a back-action on the atomic wavevector. One should note that in Eq.~(\ref{eqn:F_het}) no asumption has been done on the atom-light coupling strength. An analytical solution is obtained in the weak coupling regime and completely describes the atomic state evolution. Remarkably, two regimes occur during the wavefunction collapse: at short time spin-squeezed states are prepared whereas it converges to Dicke states at long time. 

Finally, experimental considerations to implement the detection scheme have been presented, pointing out the main advantages and limitations of the method as well as the role played by the atomic decay and its influence on the squeezing factor. The requirement of an active stabilisation of the separated geometrical paths used in standard interferometric methods, such as the Mach-Zehnder configuration, is replaced by the need for a stable RF field. Moreover, the heterodyne detection allows to reach the shot-noise of a weak beam by mixing it with a stronger one.

This detection method could be used to generate other kinds of non-classical states such as Schr{\"o}dinger cats or NOON states. Such states could be employed in quantum enhanced metrology to approach the Heisenberg limit sensitivity \cite{lee02}. Moreover, sequential measurements allow to prepare almost deterministic states by implementing a quantum feedback loop \cite{wiseman94, thomsen02, stockton04, geremia03}, and can be exploited for quantum error correction \cite{ahn03}.

\section*{Acknowledgements}

We acknowledge Isabelle Bouchoule, Carlos Garrido Alzar and Luca Pezz{\'e} for useful discussions. This work was supported by IFRAF, DGA, the European Union (with STREP program FINAQS), and ESF (EUROQUASAR program). A. B. acknowledges
support from EU under an IEF Grant.

\appendix

\section{\label{app:sp_bs}The phase modulator as a single-photon spectral beamsplitter}

The electro-optic modulation relies on the Pockels effect, a second order non-linear interaction between an optical and a quasi-static, radio-frequency (RF) field. The Hamiltonian associated with this process is:
\begin{equation}
H_{P} = i \hbar g \left( a_{m}^{\dagger} a_{0}^{\dagger} a_{1} - a_{m} a_{0} a_{1}^{\dagger}\right),
\end{equation}
where $a_{m}$, $a_{0}$ and $a_{1}$ are the annihilition operators for the RF field, the input field and the generated field, respectively and $g$ is a coupling constant (depending on the non-linear crystal, the field frequencies, \dots). This Hamiltonian is the result of two processes: a down conversion from an optical to a quasi-static electric field, that is an optical rectification, and a frequency sum with a quasi-static electric field, which is the Pockels effect.

\begin{figure}[!h]
\begin{center}
\includegraphics[width=8cm,keepaspectratio]{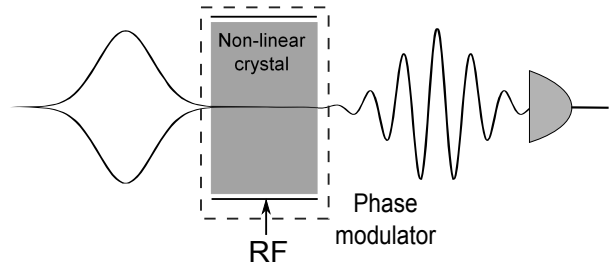}
\caption{Quantum beats of the single-photon wavefunction generated by a phase modulator.}
\label{fig:quantum_beats}
\end{center}
\end{figure}

Let $t$ be the propagation time of the optical wave in the non-linear crystal. The evolution of an incident state $\left| \psi_{\rm in} \right\rangle$ through the crystal in the weak coupling limit ($gt \ll 1$) gives the following output state:
\begin{eqnarray}
\left| \psi_{\mathrm{out}} \right\rangle & = & e^{-i H_{P} t / \hbar} \left| \psi_{\mathrm{in}} \right\rangle \\
& \sim & \left[ \mathds{1} + gt \left( a_{m}^{\dagger} a_{0}^{\dagger} a_{1} - a_{m} a_{0} a_{1}^{\dagger}\right) \right] \left| \psi_{\mathrm{in}} \right\rangle.
\end{eqnarray}
Considering a coherent RF field and a single-photon optical state in input $\left| \psi_{\mathrm{in}} \right\rangle = \left| \alpha_{m} \right\rangle \otimes \left| 1_{0}, 0_{1} \right\rangle$, the output field is:
\begin{equation}
\left| \psi_{\mathrm{out}} \right\rangle = \left| \alpha_{m} \right\rangle \otimes \left( \left| 1_{0}, 0_{1} \right\rangle + gt \alpha_{m} \left| 0_{0}, 1_{1} \right\rangle \right).
\end{equation}
The phase modulator thus splits the input photon wavefunction in two spectral components, whose superposition determines a beating of the photon wavefunction with itself at the modulation frequency of the RF field, as shown in Fig~\ref{fig:quantum_beats}.

\section{\label{app:deriv_F}Derivation of an analytical expression of $\left| \mathcal{F}_{N_{p}} \left( n \right) \right|^{2}$ in the weak coupling limit}

The general stochastic measurement process described by Eq.~(\ref{eqn:F_het}) is non-Markovian. In the weak coupling limit ($\phi N_{\rm at} \ll 1$) the process becomes Markovian \cite{garZol} but is not stationary. Using again the weak coupling approximation we show that the overall process can be split into a sequence of stationary sub-processes, which brings to an analytical expression of $\left| \mathcal{F}_{N_{p}} \left( n \right) \right|^{2}$.

\subsection{Evaluation of the probability $P \left( \varphi \right)$}

In the weak coupling regime the phase shift determined by the whole atomic sample is small, then $\phi n \ll 1$ for $-N_{\rm at}/2 \leq n \leq N_{\rm at}/2$. We can expand Eq.~(\ref{eqn:P_varphi}) to the first order in $\phi n$ around 
 a centered probability distribution $P_{0} \left( \varphi \right) = \left( 1 + \mathcal{C} \cos \varphi \right)/2 \pi$. We find $P\left( \varphi \right) = P_{0} \left( \varphi \right) + \delta P \left[ \delta \widetilde{\varphi} \right] \left( \varphi \right)$, where $\delta P \left[ \delta \widetilde{\varphi} \right] \left( \varphi \right) = - \mathcal{C} \sin \varphi \; \delta \widetilde{\varphi}/2 \pi$ and $\delta \widetilde{\varphi} = - \phi \left\langle J_{z} \right\rangle$ is a stochastic parameter depending on the followed trajectory. The chain of stochastic variables $\left\{ \widetilde{\varphi}_{k} \right\}_{1 \leq k \leq N_{p}}$ is reduced to a single parameter $\delta \widetilde{\varphi}_{N_{p}}$ which proves that the process is Markovian. $P_{0} \left( \varphi \right)$ is the deterministic contribution for a zero phase detection, whereas $\delta P \left[ \delta \widetilde{\varphi} \right] \left( \varphi \right)$ describes the atomic phase-shift.

\subsection{Decomposition into stationary sub-processes}

To study the evolution of the atomic state, we split the measurement into sequences of $N_{t}$ photon detections, each inducing a negligible evolution of the atomic wavefunction (the sub-processes are stationary). A small signal-to-noise ratio is required for each sequence ($\phi^{2} N_{\mathrm{at}} N_{t} \ll 1$). In the weak coupling limit the number of photons in each sequence can be large since $\phi^{2} N_{\mathrm{at}} \ll 1$.

The detection probability at the end of the $j$-th sequence of $N_{t}$ measurements is $P_{j} \left( \varphi \right) = P_{0} \left( \varphi \right) + \delta P \left[ \delta \widetilde{\varphi}_{j} \right] \left( \varphi \right)$.
If we choose a phase measurement resolution $\pi/m$ ($m \in \mathbb{N}$ and $m \gg 1$) we find from Eq.~(\ref{eqn:F_het}):
\begin{equation}
\left| \mathcal{F}_{N_{p}} \left( n \right) \right|^{2} \propto \prod_{j=1}^{N_{p}/N_{t}} \prod_{l=-m}^{m} f_{N_{p}}^{(j,l)} \left( n \right), \label{eqn:FNp}
\end{equation} 
having defined:
\begin{equation}
f_{N_{p}}^{(j,l)} \left( n \right) = \left[ 1 + \mathcal{C} \cos \left(\phi n - \pi l/m \right) \right]^{N_{j,l}}.
\end{equation}
$N_{j,l} = \frac{\pi}{m} P_{j} \left( \frac{\pi l}{m} \right) N_{t}$ is the number of measurements giving the result $\varphi=\pi l/m$ during the $j$-th sequence. 

The number of photons has to be large in the intervals $\left[-\pi, -\pi + \pi/m \right]$ and $\left[\pi - \pi/m, \pi \right]$, where $P \left( \varphi \right)$ is close to zero. This is verified for $N_{t} \gg m^{3}$, since the following condition must be satisfied when $\mathcal{C} \sim 1$: 
\begin{equation}
N_{t} \int_{\pi - \frac{\pi}{m}}^{\pi} P_{0} \left( \varphi \right) d\varphi \sim N_{t} \left[ \frac{1 - \mathcal{C}}{2m} +  \frac{\pi^{2} \mathcal{C}}{12 m^{3}} \right] \gg 1.
\end{equation}
In that case $N_{j,l} \gg 1$ and the second order term for the Taylor expansion of $f_{N_{p}}^{(j,l)} \left( n \right)$ in $\phi n$ can be identified with that of a gaussian distribution:
\begin{equation}
f_{N_{p}}^{(j,l)} \left( n \right) \propto \exp \left[ - 2 M_{l}^{2} N_{j,l} \left(n - n_{l} \right)^{2}  \right],
\label{eqn:fNp_app}
\end{equation}
where
\begin{eqnarray}
M_{l}^{2} & = & \frac{\mathcal{C} \phi^{2}}{4} \frac{\mathcal{C} + \cos \left(\pi l/m \right)}{\left( 1+ \mathcal{C} \cos \left(\pi l/m \right) \right)^{2}}, \\
n_{l} & = & \frac{1}{\phi} \frac{1+ \mathcal{C} \cos \left(\pi l/m \right)}{\mathcal{C} + \cos \left(\pi l/m \right)} \sin \left(\pi l/m \right).
\end{eqnarray} 
By converting the discrete sum over $l$ into an integral and using Eq.~(\ref{eqn:FNp}) and Eq.~(\ref{eqn:fNp_app}), we obtain Eq.~(\ref{eqn:F_gauss}).

\section{\label{app:simul}Numerical simulations}

The numerical simulation of the quantum trajectories shown in Fig.~\ref{fig:simul_het} adopts the state of Eq.~(\ref{eqn:initial_state}) as initial one. For each photon, the probability $P_{N_{p}+1} \left( \varphi \right)$ to measure a phase $\varphi$ for the next detected photon is given by Eq.~(\ref{eqn:P_varphi}). More explicitly, we use the cumulative distribution function associated to this density of probability, which is: 
\begin{equation}
F_{N_{p}+1} \left( \varphi \right) = \int_{- \pi}^{\varphi} P_{N_{p}+1} \left( \theta \right) d\theta.
\end{equation}
By generating a random number with an uniform distribution over $\left[ 0,1 \right]$ and numerically inverting $F_{N_{p}+1}$ we get the phase detected for the ($N_{p}+1$)-th photon. The recurrence relation given by Eq.~(\ref{eqn:rec_relation}) yields the new atomic distribution. The quantum trajectory is obtained iterating the sequence.


\begin{thebibliography}{39}
\expandafter\ifx\csname natexlab\endcsname\relax\def\natexlab#1{#1}\fi
\expandafter\ifx\csname bibnamefont\endcsname\relax
  \def\bibnamefont#1{#1}\fi
\expandafter\ifx\csname bibfnamefont\endcsname\relax
  \def\bibfnamefont#1{#1}\fi
\expandafter\ifx\csname citenamefont\endcsname\relax
  \def\citenamefont#1{#1}\fi
\expandafter\ifx\csname url\endcsname\relax
  \def\url#1{\texttt{#1}}\fi
\expandafter\ifx\csname urlprefix\endcsname\relax\def\urlprefix{URL }\fi
\providecommand{\bibinfo}[2]{#2}
\providecommand{\eprint}[2][]{\url{#2}}

\bibitem[{\citenamefont{Kitagawa and Ueda}(1993)}]{kitagawa93}
\bibinfo{author}{\bibfnamefont{M.}~\bibnamefont{Kitagawa}} \bibnamefont{and}
  \bibinfo{author}{\bibfnamefont{M.}~\bibnamefont{Ueda}},
  \bibinfo{journal}{Phys.\ Rev.\ A} \textbf{\bibinfo{volume}{47}},
  \bibinfo{pages}{5138} (\bibinfo{year}{1993}).

\bibitem[{\citenamefont{Wineland et~al.}(1992)\citenamefont{Wineland,
  Bollinger, Itano, Moore, and Heinzen}}]{wineland92}
\bibinfo{author}{\bibfnamefont{D.~J.} \bibnamefont{Wineland}},
  \bibinfo{author}{\bibfnamefont{J.~J.} \bibnamefont{Bollinger}},
  \bibinfo{author}{\bibfnamefont{W.~M.} \bibnamefont{Itano}},
  \bibinfo{author}{\bibfnamefont{F.~L.} \bibnamefont{Moore}}, \bibnamefont{and}
  \bibinfo{author}{\bibfnamefont{D.~J.} \bibnamefont{Heinzen}},
  \bibinfo{journal}{Phys.\ Rev.\ A} \textbf{\bibinfo{volume}{46}},
  \bibinfo{pages}{R6797} (\bibinfo{year}{1992}).

\bibitem[{\citenamefont{Dicke}(1954)}]{dicke54}
\bibinfo{author}{\bibfnamefont{R.~H.} \bibnamefont{Dicke}},
  \bibinfo{journal}{Phys.\ Rev.} \textbf{\bibinfo{volume}{93}},
  \bibinfo{pages}{99} (\bibinfo{year}{1954}).

\bibitem[{\citenamefont{Stockton et~al.}(2004)\citenamefont{Stockton, Handel,
  and Mabuchi}}]{stockton04}
\bibinfo{author}{\bibfnamefont{J.~K.} \bibnamefont{Stockton}},
  \bibinfo{author}{\bibfnamefont{R.~V.} \bibnamefont{Handel}},
  \bibnamefont{and} \bibinfo{author}{\bibfnamefont{H.}~\bibnamefont{Mabuchi}},
  \bibinfo{journal}{Phys.\ Rev.\ A} \textbf{\bibinfo{volume}{70}},
  \bibinfo{pages}{022106} (\bibinfo{year}{2004}).

\bibitem[{\citenamefont{Thiel et~al.}(2007)\citenamefont{Thiel, von Zanthier,
  Bastin, Solano, and Agarwal}}]{thiel07}
\bibinfo{author}{\bibfnamefont{C.}~\bibnamefont{Thiel}},
  \bibinfo{author}{\bibfnamefont{J.}~\bibnamefont{von Zanthier}},
  \bibinfo{author}{\bibfnamefont{T.}~\bibnamefont{Bastin}},
  \bibinfo{author}{\bibfnamefont{E.}~\bibnamefont{Solano}}, \bibnamefont{and}
  \bibinfo{author}{\bibfnamefont{G.~S.} \bibnamefont{Agarwal}},
  \bibinfo{journal}{Phys.\ Rev.\ Lett.} \textbf{\bibinfo{volume}{99}},
  \bibinfo{pages}{193602} (\bibinfo{year}{2007}).

\bibitem[{\citenamefont{Katz et~al.}(2006)\citenamefont{Katz, Ansmann,
  Bialczak, Lucero, McDermott, Neeley, Steffen, Weig, Cleland, Martinis
  et~al.}}]{katz06}
\bibinfo{author}{\bibfnamefont{N.}~\bibnamefont{Katz}},
  \bibinfo{author}{\bibfnamefont{M.}~\bibnamefont{Ansmann}},
  \bibinfo{author}{\bibfnamefont{R.~C.} \bibnamefont{Bialczak}},
  \bibinfo{author}{\bibfnamefont{E.}~\bibnamefont{Lucero}},
  \bibinfo{author}{\bibfnamefont{R.}~\bibnamefont{McDermott}},
  \bibinfo{author}{\bibfnamefont{M.}~\bibnamefont{Neeley}},
  \bibinfo{author}{\bibfnamefont{M.}~\bibnamefont{Steffen}},
  \bibinfo{author}{\bibfnamefont{E.~M.} \bibnamefont{Weig}},
  \bibinfo{author}{\bibfnamefont{A.~N.} \bibnamefont{Cleland}},
  \bibinfo{author}{\bibfnamefont{J.~M.} \bibnamefont{Martinis}},
  \bibnamefont{et~al.}, \bibinfo{journal}{Science}
  \textbf{\bibinfo{volume}{312}}, \bibinfo{pages}{1498} (\bibinfo{year}{2006}).

\bibitem[{\citenamefont{Bouyer and Kasevich}(1997)}]{bouyer97}
\bibinfo{author}{\bibfnamefont{P.}~\bibnamefont{Bouyer}} \bibnamefont{and}
  \bibinfo{author}{\bibfnamefont{M.~A.} \bibnamefont{Kasevich}},
  \bibinfo{journal}{Phys.\ Rev.\ A} \textbf{\bibinfo{volume}{56}},
  \bibinfo{pages}{R1083} (\bibinfo{year}{1997}).

\bibitem[{\citenamefont{Appel et~al.}(2009)\citenamefont{Appel, Windpassinger,
  Oblak, Hoff, Kj{\ae}rgaard, and Polzik}}]{appel08}
\bibinfo{author}{\bibfnamefont{J.}~\bibnamefont{Appel}},
  \bibinfo{author}{\bibfnamefont{P.~J.} \bibnamefont{Windpassinger}},
  \bibinfo{author}{\bibfnamefont{D.}~\bibnamefont{Oblak}},
  \bibinfo{author}{\bibfnamefont{U.~B.} \bibnamefont{Hoff}},
  \bibinfo{author}{\bibfnamefont{N.}~\bibnamefont{Kj{\ae}rgaard}},
  \bibnamefont{and} \bibinfo{author}{\bibfnamefont{E.~S.}
  \bibnamefont{Polzik}}, \bibinfo{journal}{PNAS}
  \textbf{\bibinfo{volume}{106}}, \bibinfo{pages}{10960}
  (\bibinfo{year}{2009}).

\bibitem[{\citenamefont{Leroux et~al.}(2010{\natexlab{a}})\citenamefont{Leroux,
  Schleier-Smith, and Vuleti{\'c}}}]{leroux10}
\bibinfo{author}{\bibfnamefont{I.~D.} \bibnamefont{Leroux}},
  \bibinfo{author}{\bibfnamefont{M.~H.} \bibnamefont{Schleier-Smith}},
  \bibnamefont{and}
  \bibinfo{author}{\bibfnamefont{V.}~\bibnamefont{Vuleti{\'c}}},
  \bibinfo{journal}{Phys.\ Rev.\ Lett.} \textbf{\bibinfo{volume}{104}},
  \bibinfo{pages}{073602} (\bibinfo{year}{2010}{\natexlab{a}}).

\bibitem[{\citenamefont{Takano et~al.}(2009)\citenamefont{Takano, Fuyama,
  Namiki, and Takahashi}}]{takano09}
\bibinfo{author}{\bibfnamefont{T.}~\bibnamefont{Takano}},
  \bibinfo{author}{\bibfnamefont{M.}~\bibnamefont{Fuyama}},
  \bibinfo{author}{\bibfnamefont{R.}~\bibnamefont{Namiki}}, \bibnamefont{and}
  \bibinfo{author}{\bibfnamefont{Y.}~\bibnamefont{Takahashi}},
  \bibinfo{journal}{Phys. Rev. Lett.} \textbf{\bibinfo{volume}{102}},
  \bibinfo{pages}{033601} (\bibinfo{year}{2009}).

\bibitem[{\citenamefont{Est{\`e}ve et~al.}(2008)\citenamefont{Est{\`e}ve,
  Gross, Giovanazzi, and Oberthaler}}]{esteve08}
\bibinfo{author}{\bibfnamefont{J.}~\bibnamefont{Est{\`e}ve}},
  \bibinfo{author}{\bibfnamefont{C.}~\bibnamefont{Gross}},
  \bibinfo{author}{\bibfnamefont{S.}~\bibnamefont{Giovanazzi}},
  \bibnamefont{and} \bibinfo{author}{\bibfnamefont{M.~K.}
  \bibnamefont{Oberthaler}}, \bibinfo{journal}{Nature (London)}
  \textbf{\bibinfo{volume}{455}}, \bibinfo{pages}{1216} (\bibinfo{year}{2008}).

\bibitem[{\citenamefont{Gross et~al.}(2010)\citenamefont{Gross, Zibold, Niklas,
  Est{\`e}ve, , and Oberthaler}}]{gross10}
\bibinfo{author}{\bibfnamefont{C.}~\bibnamefont{Gross}},
  \bibinfo{author}{\bibfnamefont{T.}~\bibnamefont{Zibold}},
  \bibinfo{author}{\bibfnamefont{E.}~\bibnamefont{Niklas}},
  \bibinfo{author}{\bibfnamefont{J.}~\bibnamefont{Est{\`e}ve}}, ,
  \bibnamefont{and} \bibinfo{author}{\bibfnamefont{M.~K.}
  \bibnamefont{Oberthaler}}, \bibinfo{journal}{Nature (London)}
  \textbf{\bibinfo{volume}{464}}, \bibinfo{pages}{1165} (\bibinfo{year}{2010}).

\bibitem[{\citenamefont{Kuzmich et~al.}(1997)\citenamefont{Kuzmich, M{\o}lmer,
  and Polzik}}]{kuzmich97}
\bibinfo{author}{\bibfnamefont{A.}~\bibnamefont{Kuzmich}},
  \bibinfo{author}{\bibfnamefont{K.}~\bibnamefont{M{\o}lmer}},
  \bibnamefont{and} \bibinfo{author}{\bibfnamefont{E.~S.}
  \bibnamefont{Polzik}}, \bibinfo{journal}{Phys.\ Rev.\ Lett.}
  \textbf{\bibinfo{volume}{79}}, \bibinfo{pages}{4782} (\bibinfo{year}{1997}).

\bibitem[{\citenamefont{Bouchoule and M{\o}lmer}(2002)}]{bouchoule02}
\bibinfo{author}{\bibfnamefont{I.}~\bibnamefont{Bouchoule}} \bibnamefont{and}
  \bibinfo{author}{\bibfnamefont{K.}~\bibnamefont{M{\o}lmer}},
  \bibinfo{journal}{Phys.\ Rev.\ A} \textbf{\bibinfo{volume}{66}},
  \bibinfo{pages}{043811} (\bibinfo{year}{2002}).

\bibitem[{\citenamefont{Kuzmich et~al.}(1998)\citenamefont{Kuzmich, Bigelow,
  and Mandel}}]{kuzmich98}
\bibinfo{author}{\bibfnamefont{A.}~\bibnamefont{Kuzmich}},
  \bibinfo{author}{\bibfnamefont{N.~P.} \bibnamefont{Bigelow}},
  \bibnamefont{and} \bibinfo{author}{\bibfnamefont{L.}~\bibnamefont{Mandel}},
  \bibinfo{journal}{Europhys.\ Lett.} \textbf{\bibinfo{volume}{42}},
  \bibinfo{pages}{481} (\bibinfo{year}{1998}).

\bibitem[{\citenamefont{Kuzmich et~al.}(2000)\citenamefont{Kuzmich, Mandel, and
  Bigelow}}]{kuzmich00}
\bibinfo{author}{\bibfnamefont{A.}~\bibnamefont{Kuzmich}},
  \bibinfo{author}{\bibfnamefont{L.}~\bibnamefont{Mandel}}, \bibnamefont{and}
  \bibinfo{author}{\bibfnamefont{N.~P.} \bibnamefont{Bigelow}},
  \bibinfo{journal}{Phys.\ Rev.\ Lett.} \textbf{\bibinfo{volume}{85}},
  \bibinfo{pages}{1594} (\bibinfo{year}{2000}).

\bibitem[{\citenamefont{Nielsen et~al.}(2009)\citenamefont{Nielsen, Poulsen,
  Negretti, and M{\o}lmer}}]{nielsen09}
\bibinfo{author}{\bibfnamefont{A.~E.~B.} \bibnamefont{Nielsen}},
  \bibinfo{author}{\bibfnamefont{U.~V.} \bibnamefont{Poulsen}},
  \bibinfo{author}{\bibfnamefont{A.}~\bibnamefont{Negretti}}, \bibnamefont{and}
  \bibinfo{author}{\bibfnamefont{K.}~\bibnamefont{M{\o}lmer}},
  \bibinfo{journal}{Phys.\ Rev.\ A} \textbf{\bibinfo{volume}{79}},
  \bibinfo{pages}{023841} (\bibinfo{year}{2009}).

\bibitem[{\citenamefont{Saffman et~al.}(2009)\citenamefont{Saffman, Oblak,
  Appel, and Polzik}}]{saffman09}
\bibinfo{author}{\bibfnamefont{M.}~\bibnamefont{Saffman}},
  \bibinfo{author}{\bibfnamefont{D.}~\bibnamefont{Oblak}},
  \bibinfo{author}{\bibfnamefont{J.}~\bibnamefont{Appel}}, \bibnamefont{and}
  \bibinfo{author}{\bibfnamefont{E.~S.} \bibnamefont{Polzik}},
  \bibinfo{journal}{Phys.\ Rev.\ A} \textbf{\bibinfo{volume}{79}},
  \bibinfo{pages}{023831} (\bibinfo{year}{2009}).

\bibitem[{\citenamefont{Schleier-Smith
  et~al.}(2010)\citenamefont{Schleier-Smith, Leroux, and
  Vuleti{\'c}}}]{schleier10}
\bibinfo{author}{\bibfnamefont{M.~H.} \bibnamefont{Schleier-Smith}},
  \bibinfo{author}{\bibfnamefont{I.~D.} \bibnamefont{Leroux}},
  \bibnamefont{and}
  \bibinfo{author}{\bibfnamefont{V.}~\bibnamefont{Vuleti{\'c}}},
  \bibinfo{journal}{Phys.\ Rev.\ Lett.} \textbf{\bibinfo{volume}{104}},
  \bibinfo{pages}{073604} (\bibinfo{year}{2010}).

\bibitem[{\citenamefont{Leroux et~al.}(2010{\natexlab{b}})\citenamefont{Leroux,
  Schleier-Smith, and Vuleti{\'c}}}]{leroux10bis}
\bibinfo{author}{\bibfnamefont{I.~D.} \bibnamefont{Leroux}},
  \bibinfo{author}{\bibfnamefont{M.~H.} \bibnamefont{Schleier-Smith}},
  \bibnamefont{and}
  \bibinfo{author}{\bibfnamefont{V.}~\bibnamefont{Vuleti{\'c}}},
  \bibinfo{journal}{Phys.\ Rev.\ Lett.} \textbf{\bibinfo{volume}{104}},
  \bibinfo{pages}{250801} (\bibinfo{year}{2010}{\natexlab{b}}).

\bibitem[{\citenamefont{Nielsen and M{\o}lmer}(2008)}]{nielsen08}
\bibinfo{author}{\bibfnamefont{A.~E.~B.} \bibnamefont{Nielsen}}
  \bibnamefont{and}
  \bibinfo{author}{\bibfnamefont{K.}~\bibnamefont{M{\o}lmer}},
  \bibinfo{journal}{Phys.\ Rev.\ A} \textbf{\bibinfo{volume}{77}},
  \bibinfo{pages}{063811} (\bibinfo{year}{2008}).

\bibitem[{\citenamefont{S{\o}rensen and M{\o}lmer}(2002)}]{sorensen02}
\bibinfo{author}{\bibfnamefont{A.~S.} \bibnamefont{S{\o}rensen}}
  \bibnamefont{and}
  \bibinfo{author}{\bibfnamefont{K.}~\bibnamefont{M{\o}lmer}},
  \bibinfo{journal}{Phys.\ Rev.\ A} \textbf{\bibinfo{volume}{66}},
  \bibinfo{pages}{022314} (\bibinfo{year}{2002}).

\bibitem[{\citenamefont{Bjorklund et~al.}(1983)\citenamefont{Bjorklund,
  Levenson, Lenth, and Ortiz}}]{bjorklund83}
\bibinfo{author}{\bibfnamefont{G.~C.} \bibnamefont{Bjorklund}},
  \bibinfo{author}{\bibfnamefont{M.~D.} \bibnamefont{Levenson}},
  \bibinfo{author}{\bibfnamefont{W.}~\bibnamefont{Lenth}}, \bibnamefont{and}
  \bibinfo{author}{\bibfnamefont{C.}~\bibnamefont{Ortiz}},
  \bibinfo{journal}{Appl. Phys. B} \textbf{\bibinfo{volume}{32}},
  \bibinfo{pages}{145} (\bibinfo{year}{1983}).

\bibitem[{\citenamefont{Savalli et~al.}(1999)\citenamefont{Savalli, Horvath,
  Featonby, Cognet, Westbrook, Westbrook, and Aspect}}]{savalli99}
\bibinfo{author}{\bibfnamefont{V.}~\bibnamefont{Savalli}},
  \bibinfo{author}{\bibfnamefont{G.~Z.~K.} \bibnamefont{Horvath}},
  \bibinfo{author}{\bibfnamefont{P.~D.} \bibnamefont{Featonby}},
  \bibinfo{author}{\bibfnamefont{L.}~\bibnamefont{Cognet}},
  \bibinfo{author}{\bibfnamefont{N.}~\bibnamefont{Westbrook}},
  \bibinfo{author}{\bibfnamefont{C.~I.} \bibnamefont{Westbrook}},
  \bibnamefont{and} \bibinfo{author}{\bibfnamefont{A.}~\bibnamefont{Aspect}},
  \bibinfo{journal}{Opt.\ Lett.} \textbf{\bibinfo{volume}{24}},
  \bibinfo{pages}{1552} (\bibinfo{year}{1999}).

\bibitem[{\citenamefont{Lye et~al.}(2003)\citenamefont{Lye, Hope, and
  Close}}]{lye03}
\bibinfo{author}{\bibfnamefont{J.~E.} \bibnamefont{Lye}},
  \bibinfo{author}{\bibfnamefont{J.~J.} \bibnamefont{Hope}}, \bibnamefont{and}
  \bibinfo{author}{\bibfnamefont{J.~D.} \bibnamefont{Close}},
  \bibinfo{journal}{Phys.\ Rev.\ A} \textbf{\bibinfo{volume}{67}},
  \bibinfo{pages}{043609} (\bibinfo{year}{2003}).

\bibitem[{\citenamefont{Teper et~al.}(2008)\citenamefont{Teper, Vrijsen, Lee,
  and Kasevich}}]{teper08}
\bibinfo{author}{\bibfnamefont{I.}~\bibnamefont{Teper}},
  \bibinfo{author}{\bibfnamefont{G.}~\bibnamefont{Vrijsen}},
  \bibinfo{author}{\bibfnamefont{J.}~\bibnamefont{Lee}}, \bibnamefont{and}
  \bibinfo{author}{\bibfnamefont{M.~A.} \bibnamefont{Kasevich}},
  \bibinfo{journal}{Phys.\ Rev.\ A} \textbf{\bibinfo{volume}{78}},
  \bibinfo{pages}{051803(R)} (\bibinfo{year}{2008}).

\bibitem[{\citenamefont{Brune et~al.}(1990)\citenamefont{Brune, Haroche,
  Lefevre, Raimond, and Zagury}}]{brune90}
\bibinfo{author}{\bibfnamefont{M.}~\bibnamefont{Brune}},
  \bibinfo{author}{\bibfnamefont{S.}~\bibnamefont{Haroche}},
  \bibinfo{author}{\bibfnamefont{V.}~\bibnamefont{Lefevre}},
  \bibinfo{author}{\bibfnamefont{J.-M.} \bibnamefont{Raimond}},
  \bibnamefont{and} \bibinfo{author}{\bibfnamefont{N.}~\bibnamefont{Zagury}},
  \bibinfo{journal}{Phys.\ Rev.\ Lett.} \textbf{\bibinfo{volume}{65}},
  \bibinfo{pages}{976} (\bibinfo{year}{1990}).

\bibitem[{\citenamefont{Guerlin et~al.}(2007)\citenamefont{Guerlin, Bernu,
  Del{\'e}glise, Sayrin, Gleyzes, Kuhr, Brune, Raimond, and
  Haroche}}]{guerlin07}
\bibinfo{author}{\bibfnamefont{C.}~\bibnamefont{Guerlin}},
  \bibinfo{author}{\bibfnamefont{J.}~\bibnamefont{Bernu}},
  \bibinfo{author}{\bibfnamefont{S.}~\bibnamefont{Del{\'e}glise}},
  \bibinfo{author}{\bibfnamefont{C.}~\bibnamefont{Sayrin}},
  \bibinfo{author}{\bibfnamefont{S.}~\bibnamefont{Gleyzes}},
  \bibinfo{author}{\bibfnamefont{S.}~\bibnamefont{Kuhr}},
  \bibinfo{author}{\bibfnamefont{M.}~\bibnamefont{Brune}},
  \bibinfo{author}{\bibfnamefont{J.-M.} \bibnamefont{Raimond}},
  \bibnamefont{and} \bibinfo{author}{\bibfnamefont{S.}~\bibnamefont{Haroche}},
  \bibinfo{journal}{Nature (London)} \textbf{\bibinfo{volume}{448}},
  \bibinfo{pages}{889} (\bibinfo{year}{2007}).

\bibitem[{\citenamefont{Zare}(1988)}]{zare}
\bibinfo{author}{\bibfnamefont{R.~N.} \bibnamefont{Zare}},
  \emph{\bibinfo{title}{Angular momemtum}} (\bibinfo{publisher}{John Wiley and
  Sons, New York}, \bibinfo{year}{1988}).

\bibitem[{\citenamefont{Adler et~al.}(2001)\citenamefont{Adler, Brody, Brun,
  and Hughston}}]{adler01}
\bibinfo{author}{\bibfnamefont{S.~L.} \bibnamefont{Adler}},
  \bibinfo{author}{\bibfnamefont{D.~C.} \bibnamefont{Brody}},
  \bibinfo{author}{\bibfnamefont{T.~A.} \bibnamefont{Brun}}, \bibnamefont{and}
  \bibinfo{author}{\bibfnamefont{L.~P.} \bibnamefont{Hughston}},
  \bibinfo{journal}{J.\ Phys.\ A} \textbf{\bibinfo{volume}{34}},
  \bibinfo{pages}{8795} (\bibinfo{year}{2001}).

\bibitem[{\citenamefont{Madsen and M{\o}lmer}(2004)}]{madsen04}
\bibinfo{author}{\bibfnamefont{L.~B.} \bibnamefont{Madsen}} \bibnamefont{and}
  \bibinfo{author}{\bibfnamefont{K.}~\bibnamefont{M{\o}lmer}},
  \bibinfo{journal}{Phys.\ Rev.\ A} \textbf{\bibinfo{volume}{70}},
  \bibinfo{pages}{052324} (\bibinfo{year}{2004}).

\bibitem[{\citenamefont{de~Echaniz et~al.}(2005)\citenamefont{de~Echaniz,
  Mitchell, Kubasik, Koschorreck, Crepaz, Eschner, and Polzik}}]{echaniz05}
\bibinfo{author}{\bibfnamefont{S.~R.} \bibnamefont{de~Echaniz}},
  \bibinfo{author}{\bibfnamefont{M.~W.} \bibnamefont{Mitchell}},
  \bibinfo{author}{\bibfnamefont{M.}~\bibnamefont{Kubasik}},
  \bibinfo{author}{\bibfnamefont{M.}~\bibnamefont{Koschorreck}},
  \bibinfo{author}{\bibfnamefont{H.}~\bibnamefont{Crepaz}},
  \bibinfo{author}{\bibfnamefont{J.}~\bibnamefont{Eschner}}, \bibnamefont{and}
  \bibinfo{author}{\bibfnamefont{E.~S.} \bibnamefont{Polzik}},
  \bibinfo{journal}{J. Opt. B: Quantum Semiclass. Opt.}
  \textbf{\bibinfo{volume}{7}}, \bibinfo{pages}{S548} (\bibinfo{year}{2005}).

\bibitem[{\citenamefont{Oblak et~al.}(2005)\citenamefont{Oblak, Petrov, Alzar,
  Tittel, Vershovski, Mikkelsen, S{\o}rensen, and Polzik}}]{oblack05}
\bibinfo{author}{\bibfnamefont{D.}~\bibnamefont{Oblak}},
  \bibinfo{author}{\bibfnamefont{P.~G.} \bibnamefont{Petrov}},
  \bibinfo{author}{\bibfnamefont{C.~L.~G.} \bibnamefont{Alzar}},
  \bibinfo{author}{\bibfnamefont{W.}~\bibnamefont{Tittel}},
  \bibinfo{author}{\bibfnamefont{A.~K.} \bibnamefont{Vershovski}},
  \bibinfo{author}{\bibfnamefont{J.~K.} \bibnamefont{Mikkelsen}},
  \bibinfo{author}{\bibfnamefont{J.~L.} \bibnamefont{S{\o}rensen}},
  \bibnamefont{and} \bibinfo{author}{\bibfnamefont{E.~S.}
  \bibnamefont{Polzik}}, \bibinfo{journal}{Phys.\ Rev.\ A}
  \textbf{\bibinfo{volume}{71}}, \bibinfo{pages}{043807}
  (\bibinfo{year}{2005}).

\bibitem[{\citenamefont{Lee et~al.}(2002)\citenamefont{Lee, Kok, and
  Dowling}}]{lee02}
\bibinfo{author}{\bibfnamefont{H.}~\bibnamefont{Lee}},
  \bibinfo{author}{\bibfnamefont{P.}~\bibnamefont{Kok}}, \bibnamefont{and}
  \bibinfo{author}{\bibfnamefont{J.~P.} \bibnamefont{Dowling}},
  \bibinfo{journal}{J. Mod. Opt.} \textbf{\bibinfo{volume}{49}},
  \bibinfo{pages}{2325} (\bibinfo{year}{2002}).

\bibitem[{\citenamefont{Wiseman and Milburn}(1994)}]{wiseman94}
\bibinfo{author}{\bibfnamefont{H.~M.} \bibnamefont{Wiseman}} \bibnamefont{and}
  \bibinfo{author}{\bibfnamefont{G.~J.} \bibnamefont{Milburn}},
  \bibinfo{journal}{Phys.\ Rev.\ A} \textbf{\bibinfo{volume}{49}},
  \bibinfo{pages}{1350} (\bibinfo{year}{1994}).

\bibitem[{\citenamefont{Thomsen et~al.}(2002)\citenamefont{Thomsen, Mancini,
  and Wiseman}}]{thomsen02}
\bibinfo{author}{\bibfnamefont{L.~K.} \bibnamefont{Thomsen}},
  \bibinfo{author}{\bibfnamefont{S.}~\bibnamefont{Mancini}}, \bibnamefont{and}
  \bibinfo{author}{\bibfnamefont{H.~M.} \bibnamefont{Wiseman}},
  \bibinfo{journal}{Phys.\ Rev.\ A} \textbf{\bibinfo{volume}{65}},
  \bibinfo{pages}{061801(R)} (\bibinfo{year}{2002}).

\bibitem[{\citenamefont{Geremia et~al.}(2003)\citenamefont{Geremia, Stockton,
  Doherty, and Mabuchi}}]{geremia03}
\bibinfo{author}{\bibfnamefont{J.~M.} \bibnamefont{Geremia}},
  \bibinfo{author}{\bibfnamefont{J.~K.} \bibnamefont{Stockton}},
  \bibinfo{author}{\bibfnamefont{A.~C.} \bibnamefont{Doherty}},
  \bibnamefont{and} \bibinfo{author}{\bibfnamefont{H.}~\bibnamefont{Mabuchi}},
  \bibinfo{journal}{Phys.\ Rev.\ Lett.} \textbf{\bibinfo{volume}{91}},
  \bibinfo{pages}{250801} (\bibinfo{year}{2003}).

\bibitem[{\citenamefont{Ahn et~al.}(2003)\citenamefont{Ahn, Wiseman, and
  Milburn}}]{ahn03}
\bibinfo{author}{\bibfnamefont{C.}~\bibnamefont{Ahn}},
  \bibinfo{author}{\bibfnamefont{H.~M.} \bibnamefont{Wiseman}},
  \bibnamefont{and} \bibinfo{author}{\bibfnamefont{G.~J.}
  \bibnamefont{Milburn}}, \bibinfo{journal}{Phys.\ Rev.\ A}
  \textbf{\bibinfo{volume}{67}}, \bibinfo{pages}{052310}
  (\bibinfo{year}{2003}).

\bibitem[{\citenamefont{Gardiner and Zoller}(2000)}]{garZol}
\bibinfo{author}{\bibfnamefont{C.~W.} \bibnamefont{Gardiner}} \bibnamefont{and}
  \bibinfo{author}{\bibfnamefont{P.}~\bibnamefont{Zoller}},
  \emph{\bibinfo{title}{Quantum Noise}} (\bibinfo{publisher}{Springer},
  \bibinfo{year}{2000}), \bibinfo{edition}{2nd} ed.

\end{thebibliography}
\end{document}